# Centralized Dynamic State Estimation Algorithm for Detecting and Distinguishing Faults and Cyber Attacks in Power Systems

Emad Abukhousa, Syed Sohail Feroz Syed Afroz, Fahad Alsaeed, Abdulaziz Qwbaiban, and A.P. Sakis Meliopoulos
*School of Electrical and Computer Engineering, Georgia Institute of Technology,* Atlanta, GA, USA
{emadak, safroz7, falsaeed3, aqwbaiban3, sakis.m}@gatech.edu

*Abstract*—As power systems evolve with increased integration of renewable energy sources and computer-based protection and control systems, they become more vulnerable to both cyber and physical threats. This study validates a centralized Dynamic State Estimation (DSE) algorithm designed to enhance the protection of power systems, particularly focusing on microgrids with substantial renewable sources integration. The algorithm utilizes a structured hypothesis testing framework, systematically identifies and differentiates anomalies caused by cyberattacks from those resulting from physical faults. This algorithm was evaluated through four case studies: a False Data Injection Attack (FDIA) via manipulation of Current Transformer (CT) ratios, a single line-to-ground (SLG) fault, and two combined scenarios involving both anomalies. Results from real-time simulations demonstrate that the algorithm effectively distinguishes between cyber-induced anomalies and physical faults, thereby significantly enhancing the reliability and security of energy systems. This research underscores the critical role of advanced diagnostic tools for protection and control systems against the growing prevalence of cyber-physical threats, enhancing the resilience of the grid and preventing potential blackouts by avoiding mis-operation of protective relays.

## I. INTRODUCTION

Microgrids are flexible energy systems that enhance local control over power generation and distribution, improve reliability, support the integration of renewable energy sources, and provide critical backup for essential services during emergencies. Their ability to operate independently from the main grid makes them invaluable for critical infrastructure and remote locations [1]. However, the integration of Distributed Energy Resources (DERs) and Inverter-Based Resources (IBRs) into microgrids introduces significant challenges to existing protection schemes. These systems, characterized by bidirectional power flows and variable fault current characteristics, often surpass the capabilities of traditional protective methods [2].

Additionally, the increasing reliance on cyber-physical systems within microgrids exposes them to a spectrum of cybersecurity threats. Cyber threats, including FDIAs, Distributed Denial of Service (DDoS) attacks, and man-in-the-middle (MiTM) attacks, can severely disrupt critical system operations. While FDIAs and MiTM attacks may alter or intercept critical system measurements, DDoS attacks primarily overwhelm the network, rendering communication and control systems unresponsive and compromising the integrity and reliability of protection schemes [3]-[7].

In addition to the cyber threats outlined, microgrids are also vulnerable to traditional electrical faults, such as line-to-line and line-to-ground faults, among others. These physical faults present significant challenges on their own; however, the situation becomes even more complex when cyberattacks mimic these faults. For instance, cyber attackers may generate fake signals or inject false data, leading to the mis-operation of protection schemes, either falsely triggering necessary interventions or failing to activate them when needed. This scenario underscores the critical importance of distinguishing between actual faults and cyber-induced anomalies. Accurately identifying the source of the disturbance; whether it is a genuine fault or a cyberattack, is essential for ensuring the correct operation of protection systems, thereby safeguarding the grid against both unplanned outages and potential security breaches [2].

Recent advancements focus on enhancing fault detection and cyber resilience through innovative approaches such as DSE-based protection, which improves the reliability of microgrid operations by distinguishing between actual faults and measurement errors [1], [2], [8]. Other researchers have explored passive and active monitoring techniques to protect against data injection attacks and to ensure the integrity of microgrid communications [6], [9]-[11]. Despite these developments, a significant gap exists in the literature regarding methods that clearly differentiate between cyber-attacks and conventional faults in microgrids incorporating IBRs and DERs. This differentiation is crucial for reliable operations, as the integration of these resources can obscure the origins of disturbances, complicating effective microgrid security management [1]. This gap underscores the need for targeted research to develop sophisticated diagnostic tools capable of discerning cyber-induced and fault-induced anomalies in advanced microgrid configurations.

The primary objective of this study is to evaluate the effectiveness of a centralized DSE algorithm designed to detect and differentiate cyber-attacks and electrical faults in microgrids. Building on the foundational work of researchers [1], [2], [8], this research seeks to implement a robust DSE framework integrated with a hypothesis testing algorithm to enhance the accuracy of anomaly detection within microgrid systems. The research conducts a series of targeted case studies to validate the proposed scheme, testing the algorithm under various disturbance conditions: a FDIA affecting CT ratios, a single line-to-ground fault, and a scenario combining both disturbances. The findings confirm that the algorithm effectively identifies and categorizes these disturbances, meeting the study's primary goal. Importantly, the algorithm demonstrates a strong capability to distinguish between cyber-induced anomalies and physical faults, a critical function that helps prevent mis-operations of protection relays that could lead to unnecessary power outages or system failures.

## II. DYNAMIC STATE ESTIMATION-BASED PROTECTION

Dynamic state estimation-based protection offers a "*setting-less*" approach that effectively monitors protection zones without relying on predefined configurations, thus avoiding common issues associated with setting-based relays [13]. By aligning actual measurements with a dynamic model, this method extends beyond traditional differential protection to encompass all relevant physical laws, ensuring resilience against variations in sources, faults, and loads, particularly in inverter-interfaced environments.

### A. Device/Protection zone Model

The initial step in implementing a DSE-based protection system involves developing a high-fidelity model for the protection zone, as described in [11]. This process includes constructing detailed mathematical representations for all devices, such as transmission lines, cables, inverters, energy storage systems, and distributed generation units in the protection zone. The model is developed in a standard syntax in terms of the state and controls of the protection zone, as follows:

$$\begin{aligned} i(t) &= Y_{eqx1}\mathbf{x}(t) + Y_{equ1}\mathbf{u}(t) + D_{eqxd1}\frac{d\mathbf{x}(t)}{dt} + C_{eqc1} \\ 0 &= Y_{eqx2}\mathbf{x}(t) + Y_{equ2}\mathbf{u}(t) + D_{eqxd2}\frac{d\tilde{\mathbf{x}}(t)}{dt} + C_{eqc2} \\ 0 &= Y_{eqx3}\mathbf{x}(t) + Y_{equ3}\mathbf{u}(t) + \begin{Bmatrix} \vdots \\ \mathbf{x}(t)^T\langle F^i_{eqxx3}\rangle\mathbf{x}(t) \\ \vdots \end{Bmatrix} + \begin{Bmatrix} \vdots \\ \mathbf{u}(t)^T\langle F^i_{equu3}\rangle\mathbf{u}(t) \\ \vdots \end{Bmatrix} + C_{eqc3} \end{aligned} \quad (1)$$

Where $\mathbf{x}(t)$, $\mathbf{u}(t)$ and $i(t)$ represent the states, controls, and interface currents of the model, respectively. Matrices $Y$, $D$, and $F$ are coefficient matrices, with $C$ being constant vectors. For a specific relay, the controls typically remain fixed for the duration of the fault.

### B. Measurement Model

The next step involves developing the measurement model, which requires understanding the types and locations of measurements, including actual, derived, virtual, and pseudo measurements. The general formulation of the measurement model is expressed as follows:

$$\begin{aligned} z(t) &= Y_{zx}\mathbf{x}(t) + \begin{Bmatrix} \vdots \\ \mathbf{x}^T\langle F^i_{zx}\rangle\mathbf{x} \\ \vdots \end{Bmatrix} + D_{zx}\frac{dx(t)}{dt} + C_{zx} + \eta \\ &= h(x) + \eta \end{aligned} \quad (2)$$

where $z(t)$ is the measurement vector, $h(x)$ is the deterministic component of the model that computes the expected measurement values derived from the state variables, and $\eta$ is the noise introduced by the meter.

### C. The DSE-based Algorthem

This algorithm utilizes the WLS method to obtain the most accurate parameter estimates. The optimization aims to minimize the sum of the squared differences between actual measurements and their estimates, which is formalized in the optimization problem:

$$Minimize\ J = \sum_{i=1}^{m}\left(\frac{Hx - z_i}{\sigma_i}\right)^2 \quad (3)$$

The solution to this optimization problem is given as Newton's iterative algorithm:

$$x^{v+1} = x^v - (H^TWH)^{-1}H^TW(h(x^v) - z) \quad (4)$$

Where $\sigma_i$ is the standard deviation of the measurements, $W$ is the weight matrix with the weights defined as the inverse of the variance for each measurement, and $H$ is the Jacobean matrix given as:

$$H = \frac{\partial h(x)}{\partial x}, \quad computed\ at\ x = x^v \quad (5)$$

A parameterized chi-square test is then conducted to assess the consistency between the measurements and the model. The goodness of fit between the model and the measurements is expressed as follows:

$$\zeta = \sum_{i=1}^{m}\left(\frac{h_i(x) - z_i}{\sigma_i}\right)^2 \quad (6)$$

The confidence level ($c_l$) that the measurements are consistent with the dynamic protection zone model is given as the probability:

$$Pr[\chi^2 \geq \zeta] = 1.0 - Pr[\chi^2 \leq \zeta] = 1.0 - Pr(\zeta, \nu) \quad (7)$$

Where, $\Pr(\zeta, \nu)$ is the probability function and $\nu$ is the degree of freedom. A confidence level close to 100% indicates that the measurements are consistent with the protection zone model, whereas a confidence level near 0% suggests that the measurements do not align with the model.

## III. PROTECTION SCHEME METHODOLOGY

### A. Overview of the Protection Algorithm

The flowchart in Fig. 1 provides an overview of the steps involved in the proposed microgrid protection algorithm. This algorithm was initially developed by Albinali and Meliopoulos [8] and has since been refined by Vasios [1, 2]. It is designed to detect and isolate faults while distinguishing between genuine faults and cyberattacks that lead to erroneous measurements. This study further enhances the algorithm by integrating a hypothesis testing module with the centralized DSE module, resulting in a unified central algorithm.

The algorithm begins with initialization, followed by the collection of individual device models to define their expected behaviors under normal conditions. A model of the microgrid is constructed as a baseline for anomaly detection, after which a counter is initialized to control the data collection and analysis iterations. Current and voltage measurements from the microgrid devices are then gathered to form an initial measurement model. Once this model is established, the DSE is conducted to analyze the current state of the microgrid based on the latest data. The algorithm checks for convergence; if the DSE does not converge, necessary updates are made to the measurements and model. Upon convergence, the algorithm assesses the confidence levels, indicating normal operation if above a predetermined threshold (80% in this study). Conversely, if the confidence level falls below the threshold, the hypothesis testing module is triggered to determine whether the anomaly stems from a fault or a cyber-attack. Lastly, to prevent excessive computation, the algorithm stops and return if the iteration counter exceeds a specified limit, denoted as N.

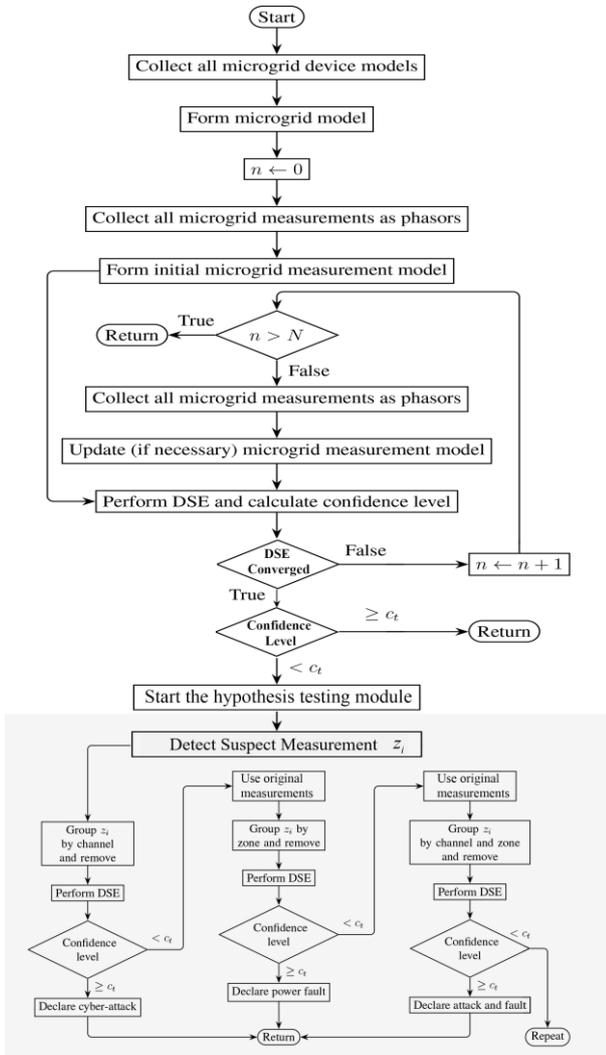

Fig. 1. Overview of the protection scheme

*B. Hypothesis testing module*

The hypothesis testing module is crucial in determining the underlying causes of anomalies in microgrid measurements, whether they arise from cyberattacks, physical faults, or both. Here's a condensed overview of the testing approach:

1) *Cyberattack Hypothesis*: This hypothesis considers a cyberattack affecting the measurement channel, causing data deviations. The suspected channel is isolated, and DSE is performed. If the confidence level after this exceeds a set threshold, it confirms a cyberattack, prompting a system declaration of a cyber incident.

2) *Fault Hypothesis:* If results from the initial test are inconclusive, the hypothesis shifts to a potential physical fault within the suspect measurement's protection zone. The zone is then isolated, and DSE is repeated. A high confidence level indicates a localized fault, leading to a declaration of a power fault.

3) *Combined Cyberattack and Fault Hypothesis:* When neither a cyberattack nor a physical fault can be distinctly identified, this hypothesis tests for the simultaneous occurrence of both. Both the channel and zone are removed for further DSE. A high confidence level supports the presence of both a fault and a cyberattack, prompting declarations accordingly. If the confidence level remains low, further investigation or alternative hypotheses may be needed.

In addition to the proposed hypotheses, it is acknowledged that other factors, such as calibration issues and instrumentation faults, can also contribute to anomalies in microgrid measurements. These issues are currently under investigation and will be integrated into future studies to enhance the comprehensiveness and accuracy of the proposed detection and classification framework.

## IV. VALIDATION THROUGH MICROGRID CASE STUDIES

The proposed protection scheme was evaluated using a microgrid test system illustrated in Fig. 2. This system, linked to the distribution network at Bus 2, includes multiple components: single-phase and three-phase loads at Buses 6 and 8, a 0.05 MVA microgrid PV system at Bus 7, and a 0.03 MVA Battery Energy Storage System (BESS) at Bus 8. The microgrid also features various lengths of cables, organized into protection zones isolated by circuit breakers to enhance reliability and fault isolation. Data collection for the protection scheme was performed using 10 merging units, which provided a combination of four current and three voltage readings each.

*A. Case Studies*

To assess the effectiveness of the proposed protection algorithm, four case studies were conducted, each simulated over a separate five-second interval, with events beginning at 2 seconds and lasting for 2 seconds. Fig. 3 illustrates the resulting phase A current waveforms for each scenario.

*1) Case Study 1: Increased CT Ratio Cyberattack on MU54*: A false data injection attack manipulated the CT ratio to 3:1, significantly increasing the amplitude of current measurements. Fig. 3a illustrates the waveform during the initial cycles of this event.

*2) Case Study 2: Single Line-to-Ground Fault:* A fault was simulated between Bus 3 and Bus 4, as shown in Fig. 3b. The line was intentionally not tripped to allow for detailed observation and analysis of the system's response.

*3) Case Study 3: Increased CT Ratio Cyberattack and Fault:* This scenario examined the system's response to a simultaneous CT cyberattack (ratio 3:1) and an SLG fault occurring within the same 2-second interval. The system's response is depicted in the zoomed-in view of Fig. 3c

*4) Case Study 4: Decreased CT Ratio Cyberattack and Fault*: This scenario evaluated a cyberattack that reduced the CT ratio to 1:5, starting at t = 2 seconds. An SLG fault was subsequently introduced at t = 2.55 seconds, lasting until t = 2.8 seconds, as shown in Fig. 3d. A different x-axis scale was used to adequately cover the event's time period.

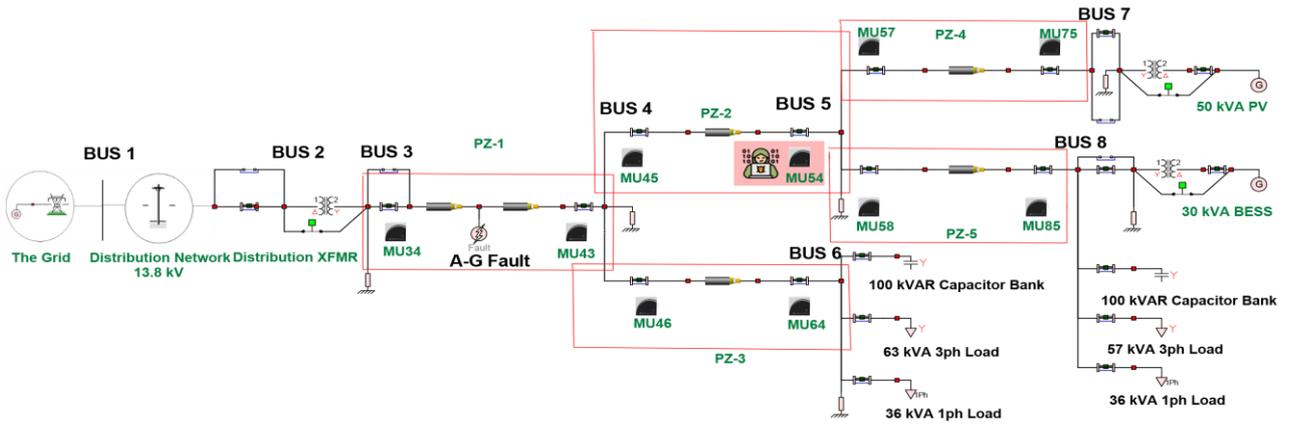

Fig. 2. Schematic diagram of the test microgrid power system

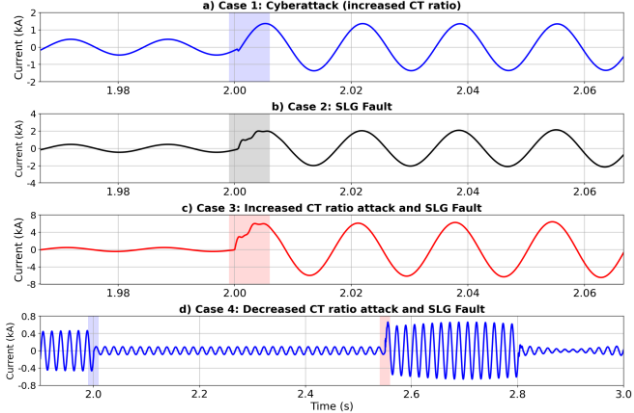

Fig. 3. Phase Current $I_A$ at MU54 for four case studies. The shaded areas in the diagrams highlight when the events started.

## B. Hypothsis testing and results

The effectiveness of the proposed algorithm was evaluated through hypothesis testing in four distinct case studies, as depicted in Fig. 4. The confidence level curve, $c_l(t)$, was analyzed before and after applying the DSE-based hypothesis testing module. In Case 1, the confidence level significantly oscillated below the predefined threshold of 80%, averaging around 40% as shown in Fig. 4a, indicating substantial data inconsistencies. The proposed algorithm responded by identifying and isolating the affected measurement channels, which were contributing to high residuals. After re-running the DSE with these channels excluded, the confidence level promptly stabilized to near 100%.

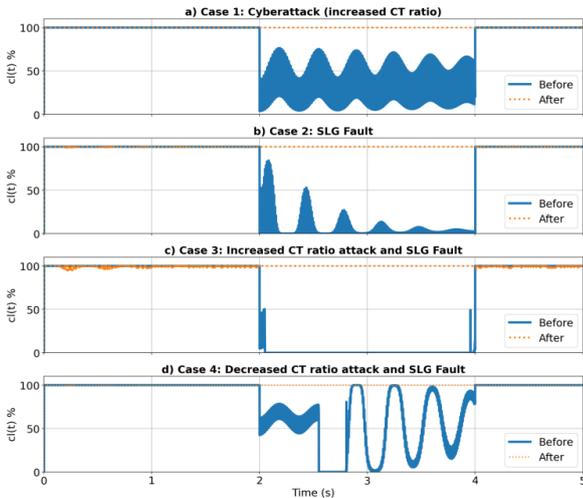

Fig. 4. Confidence levels $c_l(t)$ in present before and after applying the hypothesis module algorithm

In *Case 2*, as depicted in Fig. 4b, the SLG fault caused a sharp initial drop in confidence, followed by minor fluctuations, indicating system instability. Initial attempts to exclude only high-residual channels were insufficient to restore confidence. The algorithm refined the microgrid model by removing all measurements from the affected protection zone and adjusting related components. This restored the confidence level to near 100%, confirming a physical fault and validating the system's ability to handle localized disruptions. This process ultimately confirmed the presence of a fault, necessitating relay tripping to protect the system.

In *Case 3* (Fig. 4c), the confidence level dropped persistently below 50%, indicating simultaneous combined anomalies. Initial isolation measures, as in Case 2, were insufficient to restore confidence. The algorithm tested a third hypothesis—both a cyberattack and a fault were present—and refined the dataset by excluding suspect channels and affected protection zone data. After rerunning the DSE, the confidence level surged to nearly 100%, confirming the dual nature of the disturbances and validating the hypothesis testing approach.

Case 4 presented a complex scenario caused by a CT ratio attack that reduced the CT ratio to 5:1, causing a reduction in current measurements to 20%. This attack resulted in smaller measurements residuals compared to earlier cases and induced confidence level oscillations just above 50%, as shown in Fig. 4d. The attack was effectively detected through the chi-square test, which indicated a significant spike followed by oscillations, as highlighted in the zoomed-in view of Fig. 5b between 2 and 2.3 seconds. The proposed algorithm efficiently detected the attack, as evidenced by the confidence level oscillations depicted in both Fig. 4d and Fig. 5c and issued a decision signal—an alert—within 92 milliseconds (ms), depicted in Fig. 5d. This attack alert can be used to trigger installed protection schemes to either isolate the compromised device or replace the erroneous measurements with estimated values, thereby maintaining system integrity.

At 2.55 seconds into the scenario, an SLG fault occurred concurrently with the ongoing attack, causing the confidence level to plummet to zero. In response, the proposed algorithm issued a tripping signal within 9.4 milliseconds to isolate the affected line, as depicted in Fig. 5d, although the trip was not executed for simulation purposes. The shaded area in Fig. 5d illustrates the elapsed time from the event to the decision-making. Following the simulated cessation of the fault, the confidence level remained low due to the ongoing CT ratio attack. In response, the algorithm reapplied hypothesis testing, confirming the presence of the cyberattack and identifying the continuous confidence level oscillations caused by this persistent anomaly. Figure 4d illustrates these post-fault oscillations of $cl(t)$ curve.

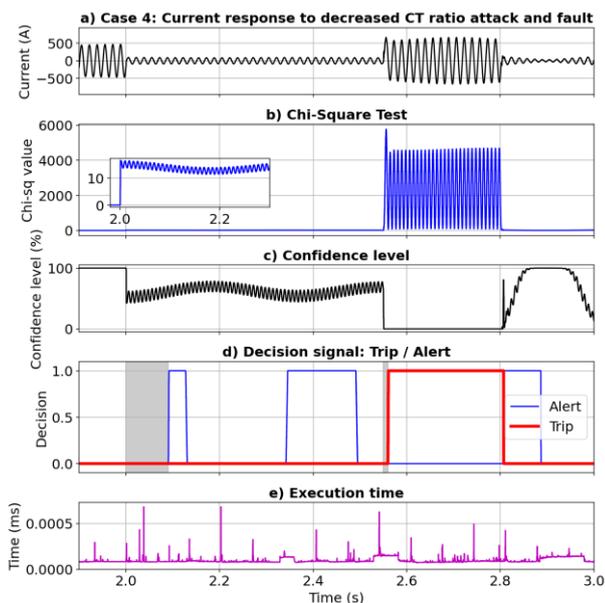

Fig. 5. Case 4: Phase IA current response, Chi-Square test results, and decision-making and execution times during attack and fault events.

*C. Summary of Timing and Calculation*

Detection and decision-making in the proposed algorithm depend on sampling rates, processing times, and user-defined delays. Using 80 samples per cycle in a 60 Hz system results in a sampling period of 208 microseconds (µs). The DSE processes two sets of samples, taking 416 µs to complete [14]. Decisions to alert or trip are made based on the area under the curve $(1-c_l(t))$ exceeding a predefined threshold $T_d$, set at 40 ms within a 100-ms moving reset window. If the area surpasses $T_d$, an anomaly is confirmed and identified, leading to a decision.

In the first three case study simulations, trip/alert signals were issued in 2.8 ms for a fault, 2.3 ms for the increased CT attack, and 2.6 ms for the combined attack and fault in Case 3. In Case 4, during the decreased CT ratio attack, the curve required 90 ms to accumulate the necessary area due to confidence level curve cl(t) oscillations above 50%. When the fault was introduced alongside the decreased CT ratio, the fault detection time under these conditions was 9.4 ms, as illustrated by the shaded areas in Fig. 5d, which indicate the time elapsed between the occurrence of the anomaly and the issuance of the decision signal. This study found that a decreased CT ratio extends the decision time by reducing the current signal, thereby compromising the system's ability to quickly detect and respond to such an attack.

*D. Threat Model Assumptions and Future Study Directions*

This study assumes a threat model in which attackers manipulate static and predefined CT ratios through the merging unit's setup interface. Recognizing the limitations of this approach, future research should explore more complex scenarios, particularly addressing FDIA that involve dynamic injections of data packets, source code alterations, or dynamic adjustments to the CT ratio to simulate gradual fluctuations in current measurements. Investigating variable CT ratio attacks is crucial, as these can dynamically change over time, mimicking normal operational fluctuations. Such attacks could complicate detection processes, delay response actions (as demonstrated in Case 4), and potentially compromise system protection mechanisms by distorting fault currents or suppressing necessary alerts.

## V. CONCLUSION AND FUTURE WORK

This study introduces a Centralized Dynamic State Estimation algorithm designed to enhance microgrid security and reliability by effectively detecting and differentiating between electrical faults and cyberattacks. Validation was achieved through four case studies: a cyberattack manipulating CT ratios, a single line-to-ground fault, and two combined scenarios of cyberattack and fault. The algorithm accurately identified these anomalies, significantly improving confidence levels after implementing the hypothesis testing framework, confirming its effectiveness in real-time environments. Notably, while the algorithm quickly detected attacks, it responded more slowly to decreased CT ratio attacks, suggesting areas for future optimization.